\documentstyle[aps,prl]{revtex}

\newcommand{\be}{\begin{equation}}
\newcommand{\ee}{\end{equation}}
\newcommand{\bea}{\begin{eqnarray}}
\newcommand{\eea}{\end{eqnarray}}
\newcommand{\beb}{\begin{eqnarray*}}
\newcommand{\eeb}{\end{eqnarray*}}

\begin{document}

\twocolumn[\hsize\textwidth\columnwidth\hsize\csname@twocolumnfalse\endcsname

\title{Theory of interlayer tunneling in bi-layer quantum Hall ferromagnets}
\author{Ady Stern$^a$, S.M. Girvin$^b$, A.H. MacDonald$^b$, and Ning Ma$^b$}
\address{
{\it (a)} Department of Condensed Matter Physics, Weizmann Institute,
Rehovot 76100, Israel\\
{\it (b)} Department of Physics, Indiana University, Bloomington,
Indiana 47405-7105}
\date{\today}
\maketitle
\begin{abstract}
Spielman et al.\ have recently observed a large zero-bias peak in the
tunnel conductance of a bi-layer system in a quantum Hall ferromagnet
state. We argue that disorder-induced topological defects in the
pseudospin order parameter limit the peak size and destroy the predicted
Josephson effect. We predict that the peak would be split and shifted by
an in-plane magnetic field in a way that maps the dispersion relation of
the ferromagnet's Goldstone mode. We also predict resonant structures in
the DC I-V characteristic under bias by an {\em ac} electric field.
\end{abstract}


\vskip2pc] 

\tighten\narrowtext

Exotic effects induced by inter-layer Coulomb interactions have made
strongly coupled bi-layer quantum Hall systems at total Landau level filling
factor $\nu =1$ the subject of numerous theoretical and experimental
studies.
\cite{reviews,fertig,macdboebinger,wenandzee,
ezawa,kyang,sternB=0,schliemann,leonleo}
When the layers are widely separated they behave as two weakly coupled $\nu
=1/2$ composite fermion metals. However, when the inter-layer distance $d$
is smaller than about twice the magnetic length $\ell$, the system
spontaneously develops interlayer phase coherence and forms
\cite{macdboebinger} a $\nu =1$ quantum Hall state.
This broken symmetry state may be described as a Bose condensate\cite
{wenandzee,ezawa} in a bosonic Chern-Simons field theory, as an easy-plane
ferromagnet\cite{macdboebinger,wenandzee,kyang,reviews} in a theory based on
a pseudospin representation for the layer degree of freedom, or as a
superfluid excitonic condensate\cite{rezayi,palacios} in a theory based on a
single-layer particle-hole transformation. We use the pseudospin language
below.

In a recent experiment, Spielman {\it et al.} \cite{spielman} observed a
qualitative change in the voltage dependence of the interlayer tunneling
current $I(V)$ upon entering the ordered state. For large $d/\ell$,
bilayer $\nu =1$ systems exhibit a pseudogap behavior: the tunneling
current is extremely small at low bias voltages. This suppression of
tunneling is attributed to the slow relaxation of charge characteristic
of the $\nu =1/2$ state in each layer. In the ordered state, Spielman
{\it et al.} discovered a strong and sharp zero bias peak in the
differential conductance $dI/dV$. It appears plausible that this peak is
related to the Josephson effect predicted by Wen and Zee\cite{wenandzee}
and by Ezawa and Iwazaki.\cite{ezawa} In contrast to the conventional
Josephson effect, however, no zero-bias supercurrent (infinite tunneling
conductance) was observed. The peak conductance, though enormously
enhanced, did not exceed $10^{-2}{e^{2}}/{h}$.

In this Letter we analyze tunneling in the bi-layer quantum Hall $\nu
=1$ state. We explain how long range density inhomogeneities introduce
topological defects (merons) into the $SU(2)$ pseudospin order
parameter. These defects carry both charge and vorticity
\cite{reviews,sinova} and constitute a dissipative environement which
turns the Josephson effect into a finite tunneling peak whose height and
width is a measure of the dynamics of the topological defects. We
predict dependences of the tunneling current on in-plane magnetic field
strength $B_{||}$, bias voltage frequency, and on the homogeneity of the
2D layers. In particular, we show that a measurement of $I(V,B_{||})$
would test the main premise of our theory, the existence of one low
energy excitation mode, and would map the dispersion relation of that
mode. Finally, we analyze the current distribution for a perfectly
homogeneous sample.

The order parameter field of the quantum Hall ferromagnet is a
pseudo-spin unit vector $\vec{m}$. When fluctuations out of the easy
plane are small, it can be parametrized by an angle $\varphi $ and the
conjugate `charge' $m_{z}$: $\vec{m}=(\cos \varphi ,\sin \varphi
,m_{z})$. In the absence of tunneling, disorder and topological defects,
the long wavelength Hamiltonian density of the $\nu =1\,$bi-layer state
is \cite{reviews,kyang,sternB=0}
\begin{equation}
H=\frac{1}{2}\rho _{s}(\nabla \varphi )^{2}+\frac{\left(
en_{0}m_{z}/2\right) ^{2}}{2\Gamma },  \label{sw-ham}
\end{equation}
where $n_{0}=\frac{1}{2\pi \ell ^{2}}$ is the average density. In
Hartree-Fock theory $\rho _{s}\sim 0.4$K and the capacitance $\Gamma $
is increased from its electrostatic value. \cite{reviews,kyang} Since
the momentum density conjugate to $\varphi $ is $p_{\varphi }=\hbar
n_{0}m_{z}/2$, the Hamiltonian (\ref{sw-ham}) has a single linearly
dispersing collective mode with velocity $u=\sqrt{\rho _{s}/\Gamma }$.
This Goldstone mode signals superfluidity for in-plane currents which
are antisymmetric in the layer index. \cite{wenandzee,ezawa,reviews}
Taking proper account of the significant exchange enhancement of $\Gamma
$ yields $u\sim 0.1{e^{2}}/{\hbar \epsilon }$. Eq.~(\ref{sw-ham}) is
valid only in the limit $q\longrightarrow 0$. Away from that limit this
collective mode has a more complicated dispersion, denoted by $\omega
_{q},$ which shows roton effects. \cite{fertig,macdboebinger} It is this
dispersion curve that may be extracted from a measurement of
$I(V,B_{||})$.

The inter-layer tunneling operators are
\begin{equation}
T_{\pm }=-\int d^{2}{r}\,\lambda (\vec{r})e^{\pm i\varphi (\vec{r})}e^{\pm
iQ_{{\rm B}}x},  \label{eq:tunnel}
\end{equation}
where the $\pm $ sign refers to the direction of tunneling, $Q_{{\rm
B}}= \frac{edB_{\parallel }}{\hbar c}$ is a characteristic wave vector
introduced \cite{reviews} by the magnetic field $B_{||}$ (we choose the
gauge $\vec{A}_{\parallel }=xB_{\parallel }\hat{z}$). The quantity
$\lambda =\frac{1}{ 8\pi \ell ^{2}}\Delta _{{\rm SAS}}$ is proportional
to the tunneling amplitude and may vary with position due to disorder in
the tunnel barrier. Here we do not discuss this source of disorder
since, on its own, it can not destroy the Josephson effect. As in a
Josephson junction, the tunneling term in the Hamiltonian is
$T_{+}+T_{-},$ while the tunneling current operator is
$ie(T_{+}-T_{-})/\hbar$. The striking similarity of the expressions
above to their counterparts in superconducting Josephson junctions make
it clear that a calculation of the tunneling conductance under Eqs.
(\ref{sw-ham}) and (\ref {eq:tunnel}) leads to a Josephson effect, in
contrast to the experiment. We now explain the way that disorder
destroys the Josephson effect in the present system.

In the $\nu =1$ bi-layer system, a deviation of the total density from
$\nu =1 $ introduces topological defects (merons) into the order
parameter vector $\vec{m}$. In terms of bosonic Chern-Simons theory,
this statement is a consequence of the residual magnetic field left,
away from $\nu =1,$ after the external and Hartree-Chern-Simons magnetic
fields almost cancel one another. This residual field introduces
vortices into the bosonic order parameters of the two layers. In the
language of a quantum Hall ferromagnet, this observation is a
consequence \cite{reviews,sternB=0} of the coupling of the symmetric
density to the order parameter $\vec{m}$. The symmetric part of the
density is constrained to satisfy $n(\vec{r})-n_{0}=\frac{1}{8\pi
}\epsilon _{ab}\epsilon _{\mu \nu \kappa }m_{\mu }\partial _{a}m_{\nu
}\partial _{b}m_{\kappa }=\nabla \cdot \left( \frac{m_{z}}{8\pi
}\hat{z}\times \nabla \varphi \right) -\frac{m_{z}}{8\pi }\nabla \times
\nabla \varphi .$ The deviation from $n_{0\text{ }}$ is then composed of
a charge density carried by an electric dipole field $\frac{m_{z}}{8\pi
}\hat{z}\times \nabla \varphi ,$ and by a charge density attached to
topological defects in $\vec{m}$. The latter are merons of four types,
carrying a charge of $\pm \frac{e}{2}$, and characterized by their
vorticity (the sign of $\nabla \times \nabla \varphi $ at the core) and
the layer in which their charge resides ($m_{z}$ at the core). Merons
interact coulombically due to their charge, and by a logarithmic
interaction due to their vorticity. Below the Kosterlitz-Thouless
temperature $T_{\rm KT}\sim\rho_s$, merons are bound in pairs of
opposite vorticity to avoid the logarithmically diverging energy
penalty.

In realistic samples there are long range density fluctuations whose
relative magnitude is estimated \cite{efros} to be 4\%. Thus, the
typical distance between the disorder-induced meron pairs is $\sim
12\ell $. The separation between the two merons that constitute a pair
is estimated to be $\sim 6\ell ,$ \cite{reviews} comparable to the
spacing among different pairs. This estimate is obtained by balancing
the Coulomb repulsion and logarithmic attraction. Thus, the $\nu =1$
bi-layer sample studied in Ref.~\onlinecite{spielman} is analogous to a
superconducting junction with random magnetic flux that introduces many
vortices in the two superconductors. Meron pairs may carry a charge $\pm
e$ (distributed between the two layers) or be charge neutral. The
charged pairs affect the longitudinal resistivity to the flow of
symmetric current. In the sample of Spielman et al.\ this resistivity is
large ($\sim 1 {\rm k}\Omega$), indicating that the charged vortex pairs
are highly mobile. Furthermore, the dissipation is not frozen out at the
lowest attainable temperatures indicating that these objects are
disorder- rather than thermally-induced. Tunneling in this system is
then strongly influenced by these merons, in a way discussed below. The
meron pairs do not however destroy the antisymmetric superfluid mode
unless they become unbound.

Appealing to the experimental observation that there is no DC Josephson
effect (i.e., current linear in the tunneling amplitude) we may use
Fermi's Golden Rule to calculate the tunneling current perturbatively.
For a sample of size $L^{2}$
\begin{equation}
I(V)=\frac{2\pi e\lambda ^{2}L^{2}}{\hbar }[S(Q_{B},eV)-S(-Q_{B},-eV)],
\label{SofQ}
\end{equation}
where $S(q,\hbar \omega ),$ the spectral density for the fluctuations of
the operator $e^{i\varphi \text{ }}$at wavevector $q$ and frequency
$\omega ,$ is proportional to the Fourier transform of $\left\langle
e^{i\varphi (r,t)\text{ }}e^{-i\varphi (0,0)\text{ }}\right\rangle $
(where angular brackets denote thermal average). Our prediction
regarding the dependence of the tunneling current on $B_{||}$ can now be
easily understood. For weak disorder, the spectral density $S(Q_{B},eV)$
is sharply peaked at
\begin{equation}
eV= \hbar \omega _{Q_{{\rm B}}}
\label{peaksplit}
\end{equation}
{\em Thus, as the parallel field is varied, the peak in the tunneling
conductance is shifted in a way that reflects the dispersion of the low
energy excitation mode.} This is precisely analogous to the
Carlson-Goldman experiment measuring the collective oscillations of the
pair field in a superconductor. \cite{carlsongoldman} An observation of
this dispersing peak would also confirm an essential ingredient of the
picture we use, namely the existence of a single branch of low energy
excitations. The parallel field allows only tunneling between states
that differ by a momentum $Q_{{\rm B}}$. Energy conservation requires
the energy of these states to differ by $eV$. When there is just one low
energy excitation branch, there is only one value of the voltage where
both these conditions are fulfilled. This is not the case for a fermi
liquid (for $Q_B\ne 0$).

To begin our analysis of the effect of merons on the tunneling
conductance we make the simplifying ansatz that the order parameter
phase can be separated into the sum a vortex part $\varphi _{m}$ and an
independent spinwave part $\varphi $ with the former obeying
$G_{m}(r,t)\equiv \langle e^{i\varphi _{m}(\vec{r},t)}e^{-i\varphi
_{m}(\vec{0},0)}\rangle =\exp (- \frac{r^{2}}{2\xi ^{2}}-\frac{t}{\tau
_{\varphi }})$. The gaussian form for the spatial dependence is chosen
for algebraic convenience. Applying this ansatz to
Eqs.(\ref{sw-ham}-\ref{SofQ}) yields
\begin{eqnarray}
I(V,B_{||}) &=&\frac{4e\lambda ^{2}L^{2}}{\hbar ^{2}}\int_{0}^{\infty
}dt\int d^2{r}\,G_{m}(r,t)e^{-\frac{1}{2}D(r,t)}\nonumber \\
&&\sin \frac{C(r,t)}{2}\cos Q_{{\rm B}}x\sin \frac{eVt}{\hbar }
  \label{summary3}
\end{eqnarray}
with a `Debye-Waller factor' $\exp(-{D(r,t)}/{2})$, where
\begin{eqnarray}
&&D(r,t)\equiv \langle \lbrack \varphi (\vec{r},t)-\varphi (\vec{0}
,0)]^{2}\rangle   \nonumber \\
&=&\sum_{q}\frac{\hbar u}{L^{2}\rho _{s}q}\left[ 1-\cos (\vec{q}\cdot
 \vec{r
})\cos (uqt)\right] \coth \frac{\hbar uq}{2T}  \label{correlator}
\end{eqnarray}
(we set $k_{\rm B}=1$ throughout) and a commutator term
\begin{eqnarray}
C(r,t) &\equiv &i[\varphi (\vec{r},t),\varphi (\vec{0},0)]  \nonumber \\
&\approx &\frac{\hbar }{2\pi \rho _{s}}\theta (ut-r)\left[ t^{2}-\left(
\frac{r}{u}\right) ^{2}\right] ^{-1/2}
\end{eqnarray}
which is independent of temperature, and limits the $r-$integral in (\ref
{summary3}) to a ``light-cone'' of $r<ut$. All correlators are evaluated in
the absence of tunneling. We rely on the global $U(1)$ symmetry and the
freedom to renormalize $\xi $ and $\tau _{\varphi }$ to partially justify
the simplifying approximation of neglecting all disorder in the spinwave
hamiltonian.

As long as $2\pi \rho _{s}\gg \hbar /\tau _{\varphi }\gg T$ we
can expand Eq.(\ref{summary3}) to first order \cite{approx} in $C$ and
approximate $D$ by its zero temperature value. In this limit the current
becomes
\begin{eqnarray}
&&I(V,B_{||})=\frac{e}{h}\frac{\xi ^{2}\lambda^2 L^{2}
}{4\Gamma }e^{-\frac{D}{2}}\int d^{2}p\,e^{-|\vec p-\vec{Q}_{B}|^{2}\xi
^{2}/2}\frac{\hbar}{\omega _{{p}}}  \nonumber \\
&&\left\{ \frac{\delta _{\varphi }}{(eV-\hbar \omega _{{p}})^{2}+(\delta
_{\varphi })^{2}}-\frac{\delta _{\varphi }}{(eV+\hbar \omega _{{p}
})^{2}+(\delta _{\varphi })^{2}}\right\}   \label{current-complicated}
\end{eqnarray}
where $\delta _{\varphi }\equiv \hbar /\tau _{\varphi }.$ For large $\tau
_{\varphi },\xi $ Eq. (\ref{current-complicated}) shows a peak in the
current at the voltage corresponding to the Goldstone mode energy in
accordance
with Eq.(\ref{peaksplit}).
The effect of $\tau _{\varphi },\xi $
is to smear this peak over a range of $\hbar /\xi $ in momentum and
$\hbar
/\tau _{\varphi }$ in voltage. As long as $Q_B\xi\gg 1$ and
$uQ_B\tau _{\varphi }\gg 1,$ this smearing is insignificant.

The expression for the differential conductance simplifies considerably in
the limit $Q_{B}=0$, $\xi \ll u\tau _{\varphi }$ and $eV\ll
\frac{\hbar u}{\xi }:$
\begin{equation}
\frac{dI}{dV}=\frac{1}{8}\frac{e^{2}}{h}\frac{\xi ^{2}}{\ell ^{2}}\frac{
n_{0}L^{2}\Delta_{SAS}^{2}}{\rho _{s}}e^{-\frac{D}{2}}\frac{\delta
_{\varphi }}{(eV)^{2}+(\delta _{\varphi })^{2}}.
\label{current-simple}
\end{equation}
Interestingly, we see that when $\tau _{\varphi }=\infty ,$ i.e., when
the merons
provide a random {\it static} background phase field, a Josephson-like
singularity of $\frac{dI}{dV}$ is still present (as is the antisymmetric
superfluid property). As shown below, the
singularity is present also at finite temperature $T\ll\rho _{s}$.  Static
topological defects break translational invariance and
thus open more phase space for excitation of spin waves in the tunneling
process. However, they do not expand the degrees of freedom involved beyond
the single spin wave mode, and thus do not dephase the process enough to
destroy the zero-voltage singularity.

The temperature dependence of (\ref{summary3}) originates from the
temperature dependence of $D$ and the temperature dependence of  $\rho _{s}$
and $\tau _{\varphi }.$ Here we calculate the temperature dependence of $D.$
At zero temperature it gives the space and time independent result
$D_{0}\equiv \int_{q\ell<\sqrt{2}}d^2q\frac{\hbar}{\Gamma u q} \sim 4.8$.
At finite temperature we approximate $\coth x\approx 1+\frac{1}{x}e^{-x}$,
define dimensionless length and time variables, $\tilde{r}
\equiv \frac{rT}{\hbar u}$ and $\tilde{t}\equiv \frac{tT}{\hbar }$, and
obtain (for large $r,t,$ and $r<ut$),
\begin{eqnarray}
&&D(\tilde{r},\tilde{t})\approx D_{0}+  \nonumber \\
&&\frac{T}{2\pi \rho _{s}}\log \left| \left( \tilde{t}+i/2\right) +
\sqrt{\left( \tilde{t}+i/2\right) ^{2}-\tilde{r}^{2}}\right| ^{2}
\label{finiteTresult}
\end{eqnarray}
The temperature dependence of $D$ affects $I(V)$ then only at high
temperature $(T\gg eV)$, where we can approximate
$\tilde{t}+i/2\approx \tilde{t}$.
For  $u\tau _{\varphi }\gg\xi$ and $B_\parallel=0$,
Eq.(\ref{summary3}) reduces to
\begin{eqnarray}
I\left( V\right)  &&\sim \frac{e\lambda ^{2}L^{2}}{\pi\rho_s\hbar}
\int_{0}^{\infty }dt\int_{r<ut}d^{2}r\exp (-\frac{1}{2}\left( \frac{r}{\xi}
\right) ^{2}-\frac{t}{\tau _{\varphi }})  \nonumber \\
&&\,\frac{\left| \frac{tT}{\hbar }+\sqrt{\left( \frac{tT}{\hbar }\right)
^{2}-\left( \frac{rT}{\hbar u}\right) ^{2}}\right| ^{-\frac{T}{2\pi \rho _{s}
}}}{\sqrt{t^{2}-(r/u)^{2}}}\sin \left( \frac{eVt}{\hbar }\right)
\end{eqnarray}
Most of the contribution is then from long times, while $r$ is limited to be
smaller than $\xi$.  For a static
meron background ($\tau _{\varphi }=\infty $) we find for $eV\ll T$
\begin{equation}
\frac{dI}{dV}\propto \frac{\lambda^2\xi^2 L^2}{\rho_s^2}
\left( \frac{T}{V}\right) ^{1-\frac{T}{2\pi \rho _{s}}}
\end{equation}
which is consistent with the more complete scaling form which can be derived
in the
classical limit from the expression of Nelson and Fisher for the
dynamical structure factor of the XY model. \cite{nelsonfisher}
In the presence of a finite $\tau _{\varphi }$, the temperature dependence of
$D$ affects the tunneling current significantly only in the window
$\frac{\hbar }{\tau _{\varphi }},eV<T<T_{\rm KT}$.
In the experiment of
Spielman et al.\  the peak width is much larger than the temperature. Thus,
the observed temperature dependence probably results from the temperature
dependence of $\rho _{s}$ and $\tau _{\varphi }$ rather than $D$.

Using Eq. (\ref{current-complicated}), we can fit the width of the
conductance peak in the experiment with a phenomenological value $\delta
_{\varphi }\approx 0.75$K. This value gives $u\tau _{\varphi }\approx 11\ell
$, which is remarkably close to our estimate of $\xi $ based on the meron
pair spacing. (Sufficiently close that the Lorentzian approximation in
Eq. (\ref{current-simple}) for the peak width will be somewhat
inaccurate.)

Our naive estimate for the peak height in the experiment is too large by some
two orders of magnitude, but is highly uncertain due to the
exponential sensitivity to the ultraviolet cutoff
and the acoustic approximation used in computing the
Debye-Waller factor. In addition, the estimate $\Delta _{SAS}\approx 90\mu $K
is exponentially sensitive to the parameters in the modeling of the barrier
potential (in particular the poorly understood effective mass appropriate
for the high Al concentration in the barrier) and so might be off by a
significant
factor. \cite{efros} It might also be possible that
the superfluidity and the tunneling occur predominately in isolated
regions close to filling factor $\nu=1$ containing few vortices.  The
parasitic series transport resistance $\sim 1/\sigma_{xx}$
in this Corbino-like geometry could significantly reduce the peak height.

We now consider inter-layer tunneling under the combined effect of a time
independent $dc$ voltage and a time dependent $ac$ electric field $E\sin
\omega t$, directed perpendicular to the two layers. As long as the system
is not heated, this field can be incorporated into our calculation by
writing $T_{\pm }=-\int d\vec{r}\,\lambda e^{\pm i[\varphi (\vec{r})+\frac{eEd
}{\hbar \omega }\cos \omega t]}$. Repeating the calculation carried out
above, we find that the tunneling differential conductance $\frac{dI}{dV}(V)$
exhibits peaks at $eV=n\hbar\omega$, with $n$ an integer. This feature is
common to all tunneling systems where the $dc$ differential conductance is
strongly peaked around zero voltage (for example, a bi-layer system at zero
magnetic field). We note, however, that the quantum Hall ferromagnet is
relatively less prone to heating, due to the small longitudinal conductivity.

Finally, we discuss the current distribution in an idealized
zero-disorder and vortex-free system.
Eqs. (\ref{sw-ham}) and (\ref{eq:tunnel}) then do indeed lead to a Sine-Gordon
equation for the phase, as in a long Josephson junction. However, due to the
two-dimensionality of the problem, the critical current is not proportional
to the area of the sample. Consider a setup where the current is fed into
one layer from, say, $x=-\infty $, and taken out from the other layer at $
x=\infty $, and where tunneling is limited to the region $-\frac{L}{2}<x<
\frac{L}{2}$. Since the symmetric part of the current ($I_{sym}$) is
conserved, the boundary conditions for the Sine-Gordon equation require $
\frac{\partial \varphi }{\partial x}|_{x=\frac{L}{2}}=-\frac{\partial
\varphi }{\partial x}|_{x=-\frac{L}{2}}=I_{sym}.$ For $L\gg\xi _{J}\equiv
\sqrt{4\pi \ell ^{2}\rho _{s}/\Delta _{{\rm SAS}}}\sim 4\mu $m the
time-independent solution to the Sine-Gordon equation in the tunneling
region is
$\varphi (x)\approx 2\arccos\tanh \frac{\frac{L}{2}-|x|}{\xi _{J}}$,
tunneling takes place only
within a distance of order $\xi _{J}$ of the $x=\pm \frac{L}{2}$ lines, and
the maximal current that can tunnel is $L-$independent, and is given by $
(2e/\hbar )\rho _{s}W/\xi $ (here $W$ is the width of the current contact).
For the parameters we use this current is $\sim 4$nA/$\mu $m$\cdot W\sim
80$nA. The experimental measurement current was much smaller than this
value. Thus, the absence of a Josephson effect cannot be attributed to a
large measurement current. For the sample geometry described above and used
in the experiment, the tunnel resistance is effectively in series with the
Hall resistance. The observed tunnel resistance is however much larger, $
\sim 10^{2}h/e^{2}$, again indicating that there is no Josephson effect.

To conclude, we have attributed the lack of a Josephson effect in
tunneling measurements in a bi-layer quantum Hall $\nu =1$ state to density
inhomogeneities that introduce dynamical topological defects into the
order parameter.  The observed peak width is quantitatively consistent
with this picture.
We showed that a measurement of the tunneling $I(V)$
dependence on $B_\parallel$
would map the dispersion relation of the low energy mode of the
system, and that tunneling in the presence of an $ac$ electric field would
result in resonances at voltages corresponding to the $ac$ frequency.
Finally, we showed that even for a perfect sample where the Josephson effect
takes place, the critical current would not scale with the size of the
sample.

AS is supported by the US-Israel BSF, and the Israeli Academy of Science,
Victor Ehrlich chair, and a DIP-BMBF grant.
The Indiana group is supported by NSF DMR-9714055. The
authors are grateful to J. P. Eisenstein and I. Ussishkin
for numerous useful discussions.


\begin{references}
\bibitem{approx}  The validity of this approximation may be checked by a
change of integration variable from $r$ to $C(r,t)$ in Eq. (\ref{summary3}).
\end{references}
\end{document}